\documentclass[twocolumn,showpacs,superscriptaddress,amsmath,amssymb,nofootinbib]{revtex4}

\usepackage{graphicx}%
\usepackage{dcolumn}%

\begin{document}

\title{ARPES on HTSC: simplicity vs. complexity}

\author{A. A. Kordyuk}
\affiliation{Institute of Metal Physics of National Academy of Sciences of Ukraine, 03142 Kyiv, Ukraine}
\affiliation{Leibniz-Institut f\"ur Festk\"orper- und Werkstoffforschung Dresden, P.O.Box 270016, 01171 Dresden, Germany}

\author{S. V. Borisenko}
\affiliation{Leibniz-Institut f\"ur Festk\"orper- und Werkstoffforschung Dresden, P.O.Box 270016, 01171 Dresden, Germany}

\date{October 1, 2005}%

\begin{abstract}
A notable role in understanding of microscopic electronic properties of high temperature superconductors (HTSC) belongs to angle resolved photoemission spectroscopy (ARPES). This technique supplies a direct window into reciprocal space of solids: the momentum-energy space where quasiparticles (the electrons dressed in clouds of interactions) dwell. Any interaction in the electronic system, e.g.~superconducting pairing, leads to modification of the quasi-particle spectrum---to redistribution of the spectral weight over the momentum-energy space probed by ARPES. A continued development of the technique had an effect that the picture seen through the ARPES window became clearer and sharper until the complexity of the electronic band structure of the cuprates had been resolved. Now, in an optimal for superconductivity doping range, the cuprates much resemble a normal metal with well predicted electronic structure, though with rather strong electron-electron interaction. This principal disentanglement of the complex physics from complex structure reduced the mystery of HTSC to a tangible problem of interaction responsible for quasi-particle formation. Here we present a short overview of resent ARPES results, which, we believe, denote a way to resolve the HTSC puzzle.
\end{abstract}


\preprint{\textit{xxx}}

\maketitle

\section{Introduction to HTSC complexity}

\begin{figure}[b]
\includegraphics[width=8cm]{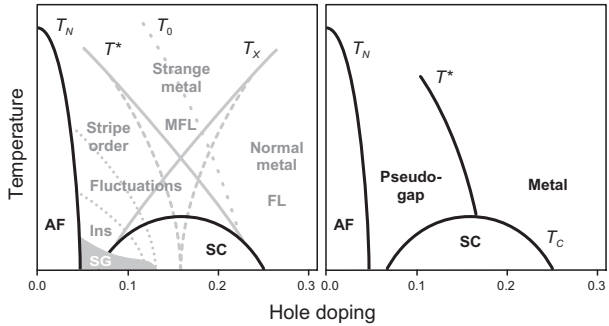}%
\caption{\label{Fig1} A schematic phase diagrams of the cuprates: generally accepted version (right) and a \textquotedblleft discussable" one (left). The antiferromagnetic phase (AF), superconducting phase (SC), as well as a crossover, $T$*, to a pseudogap region are well established \cite{1, 2}, while a number of other regions are a matter of scientific discussions \cite{3}. Here, together with the $T$* lines that confine a region of fluctuations either of superconducting \cite{4} or \textquotedblleft competing order" \cite{5} (e.g., AF, charge density waves, or spin-charge ordering), we plot a $T_x$ crossover either between marginal Fermi-liquid (MFL) and usual FL or between incoherent and coherent states of a \textquotedblleft slave" boson (see \cite{6} and references therein). Different regions at lower doping are proposed for different stripe phases \cite{7} which can result in the insulating (Ins) and spin-glass (SG) regions \cite{3, 7}. }
\end{figure}

It is real complexity of the electronic properties of the cuprates which keeps the HTSC problem open over the last 20 years. On a large scale, this complexity manifests itself in a sophisticated phase diagram, which is schematically shown in Fig.~1. In case of undoped CuO planes, the strong Coulomb repulsion between electrons indeed arranges them into anti-ferromagnetic insulating state which, upon the hole doping, evolves into a \textquotedblleft strange" metal with a sophisticated magnetic spectrum and, upon further doping, in a superconductor. On the right panel we plot such a set of phases which are mostly agreed now \cite{1, 2} while on the left panel we indicate a variety of \textquotedblleft crossover" lines which represent other suggestions \cite{3,4,5,6,7}. Living the discussion of all these crossovers beyond the scope of this paper, we note that the complexity of the phase diagram evidently is a result of an intricate evolution of the low energy electronic excitation spectrum with doping and temperature. 

The scope of this paper is to overview the recent results of the angle resolved photoemission spectroscopy (ARPES), the most direct and powerful technique that allows to look in depth on such an evolution. The aim of this review is rather ambitious: to separate the wheat from the chaff in ARPES spectra on the way to resolving the HTSC puzzle. We do it in two steps. First, we disentangle the interaction effects from the structural ones and show that the former can be described by quasiparticle self-energy. Second, we distinguish two main contributions to the self-energy, locating the key interaction which should be responsible for the superconducting pairing.

\section{Introduction to ARPES on HTSC}

In this chapter we give a brief introduction to ARPES experiment. For more details of the technique, the history of its development, as well as the majority of the obtained results, we recommend recent comprehensive reviews \cite{8, 9}.

The ARPES is an advanced development of the photoelectric effect, the theory of which has been suggested by Einstein 100 years ago. Putting a monochromatic light on a sample surface and resolving the kinetic energy of outgoing electrons one gets a photo­emission spectrometer. Resolving the angle of the photoelectrons flying out of a single crystal gives the ARPES. In modern ARPES analyzers the angle resolution up to $0.1-0.2^{\circ}$ can be achieved without any mechanical movement by using a channel-plate as a detector. For 2D solids, such as HTSC, the ARPES snapshot, $I(k, \omega)$, the image recorded by such a channelplate, is just the density distribution of the elementary electronic excitations,\footnote{A half-empiric explanation why the measured photoelectron carries information about the properties of the remaining excitation (a photohole) is based on the \textquotedblleft three step model" \cite{8}. In simple words one can say that the photoelectron gets information about interaction on the first step of photoemission breaking the coupling with other electrons.} i.e.~quasiparticles,\footnote{Here we call \textquotedblleft quasiparticles" any elementary electronic excitations which can be described in terms of Dyson self-energy \cite{10} by Eq.~(1). The excitations with a non-zero coherent component \cite{11} we will call \textquotedblleft well-defined quasiparticles".} over energy $\omega$ and momentum $k$ along a certain direction in the reciprocal space. Fig.~2 gives examples of such a distribution. Rotating the sample with respect to the analyzer one can see any such a cut of 3D momentum-energy space $I(\mathbf{k}, \omega)$, $\mathbf{k} = (k_x, k_y)$. The main region of interest is a \textquotedblleft low-energy" region of about 0.5 eV in depth from the Fermi-level, where interactions form a background for superconducting coupling.

Disregarding the effect of the energy and momentum resolutions as well as the matrix elements effect \cite{13} and the extrinsic background \cite{14, 15}, the photocurrent intensity is proportional to a one-particle spectral function multiplied by the Fermi function: $I(\mathbf{k}, \omega) = A(\mathbf{k}, \omega) f(\omega)$, where both also depend on temperature. For a non-interacting case, $A(\mathbf{k}, \omega) = \delta [\omega - \varepsilon(\mathbf{k})]$, so, the electronic spectrum is completely defined by the bare band dispersion $\varepsilon(\mathbf{k})$. Electronic interactions, which can be described in terms of the quasiparticle self-energy, $\Sigma = \Sigma' + i\Sigma''$, turn the spectral function into a renormalized form
\begin{eqnarray}\label{E1}
A(\omega, \mathbf{k}) = -\frac{1}{\pi}\frac{\Sigma''(\omega)}{(\omega - \varepsilon(\mathbf{k}) - \Sigma'(\omega))^2 + \Sigma''(\omega)^2}.
\end{eqnarray}
This is the central formula for one-particle excitation spectra analysis and it has been shown to be perfectly applicable to HTSC \cite{16}. 

In order to introduce some of ARPES vocabulary, we consider a structure of such a 3D momentum-energy space. The energy distribution curve (EDC), a unit of information of ARPES in former days, is the photocurrent intensity at certain momentum $\mathbf{k}_0$: EDC($\omega$) $= I(\mathbf{k}_0, \omega)$. The momentum distribution curve, MDC$(k) = I(k, \omega_0)$ \cite{17}, has an advantages that (i) it has a simple Lorentzian lineshape as soon as the self-energy in Eq.(1) can be considered as momentum independent, (ii) the Fermi-function and extrinsic background do not effect its lineshape (disregarding again the energy resolution effect), and (iii) a set of MDC maxima $k_m(\omega)$, by virtue of definition $\omega - \varepsilon(k_m) - \Sigma'(\omega) = 0$, trace the renormalized dispersion. In two dimensions, $I(\mathbf{k}, \omega_0)$ gives the momentum distribution map (MDM) \cite{15}, with a special case of $I(\mathbf{k}, 0)$, which is an image of the Fermi surface (FS) \cite{18}. Examples of such FS's are shown in Fig.~3 \cite{19}. Fig.~4a \cite{20} can help to navigate in the momentum space: the $\Gamma$YM triangle represents three important directions in the 1st Brillouin zone (BZ). The bare electronic structure along this triangle is shown in Fig.~4b. One can also specify two key cuts, called $\Gamma$Y and XMY, the ARPES snapshots along which are shown in Figs.~6a and 2, respectively. These two snapshots represent the \textquotedblleft nodal" and \textquotedblleft antinodal" regions\footnote{This terminology starts from the nodal direction, i.e. $(0,0)-(\pm\pi,\pm\pi)$ or $\Gamma$X(Y) directions in the BZ, along which the superconducting $d$-wave gap function has a node-changes sign. The crossing point of the nodal direction with the FS is called the nodal point (N). On the contrary, the point on the FS where the gap is believed to be maximal, or more precisely where the FS crosses the BZ boundary, is called the \textquotedblleft antinodal" point (A). Respectively, the regions around N- and A-points are called nodal and antinodal regions. Although the meaning of \textquotedblleft antinodal" region differs from paper to paper, in the most common definition it is thought as an area around the $(\pi,0)$-point which covers the nearest A-points on the FS.} and contain the essentials of the whole low energy electronic structure of the cuprates. Historically speaking, these essentials can be further reduced to a couple of heavily discussed plots shown in Fig.~5: (a) a renormalized dispersion with $\sim$ 70 meV \textquotedblleft kink", and (b) a $(\pi,0)$ EDC with so called \textquotedblleft peak-dip-hump" (PDH) structure. 

\begin{figure}[t]
\includegraphics[width=7.8cm]{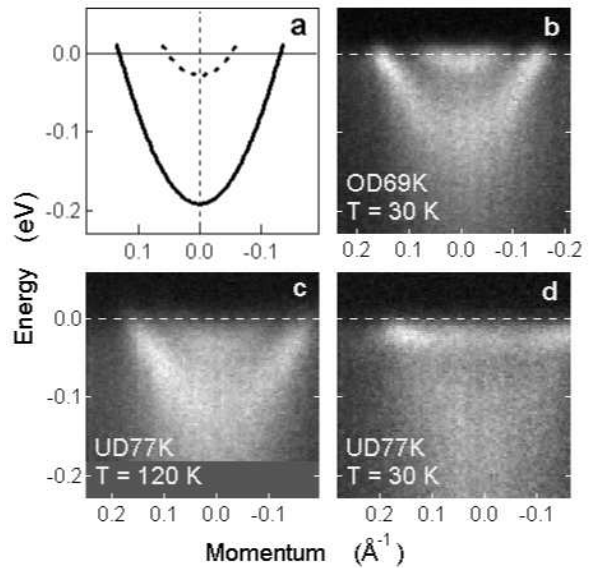}%
\caption{\label{Fig2} Electronic structure in the antinodal region (along XMY cut). Bare band structure (a) and ARPES snapshots taken at 30 K (below $T_c$) for an overdoped sample (OD, $T_c$ = 69 K, $x$ = 0.22) (b) and underdoped sample (UD, 77 K) (d), and for UD77 at 120 K (c). On panels (b) and (c) two split bands are well visible, while on panel (d) a strong depletion of the spectral can be seen at some \textquotedblleft mode" energy \cite{12}.}
\end{figure}

Before going further we want to highlight three cornerstones of our experiment (for details see \cite{12, 13, 19}). They are: (1) the precise cryo-manipulator, which allows to rotate the sample around three perpendicular axes with 0.1$^{\circ}$ precision in ultra-high vacuum (UHV); (2) the light sources (mainly synchrotrons) of a wide excitation energy range and different polarizations; and (3) high quality samples---the superstructure-free \cite{23} single crystals. To be perfectly \textquotedblleft arpesable" \cite{21}, a sample should be: (i) highly anisotropic (to be as 2D as possible, in order to eliminate the influence of the $k_z$ dispersion); (ii) easily cleavable in UHV, to reveal a perfectly flat surface; and have a surface representative of the bulk \cite{22}. The only HTSC which fits all of these is Bi-based cuprate, especially Bi(Pb)-2212, the compound with two CuO layers per unit cell, and doped with led, in order to get rid off a superstructure in the topmost BiO layer \cite{23, 24}. Therefore, here we mainly discuss the data for Bi(Pb)-2212 samples, although should note that they are in agreement with the spectra measured for La- \cite{kinks} and Y- \cite{YBCO} based samples yet of lower quality.

\begin{figure}[t]
\includegraphics[width=8cm]{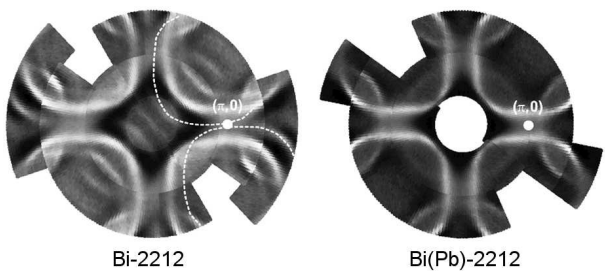}%
\caption{\label{Fig3} Fermi surfaces measured at a pristine Bi-2212 (left) and a superstructure-free lead doped Bi(Pb)-2212 (right) \cite{34}.}
\end{figure}

\begin{figure*}[t]
\includegraphics[width=4.77cm]{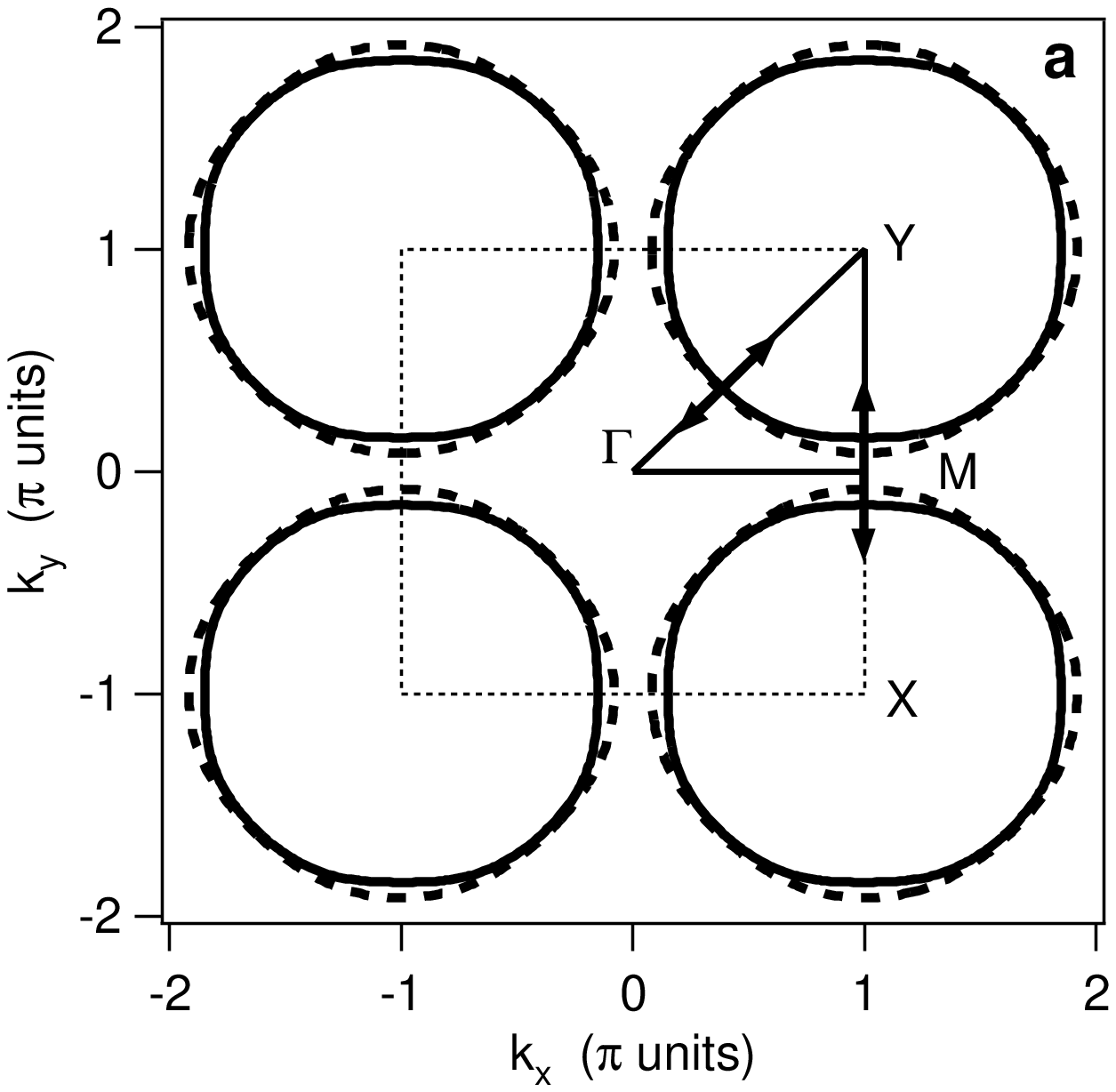}%
\includegraphics[width=9.658cm]{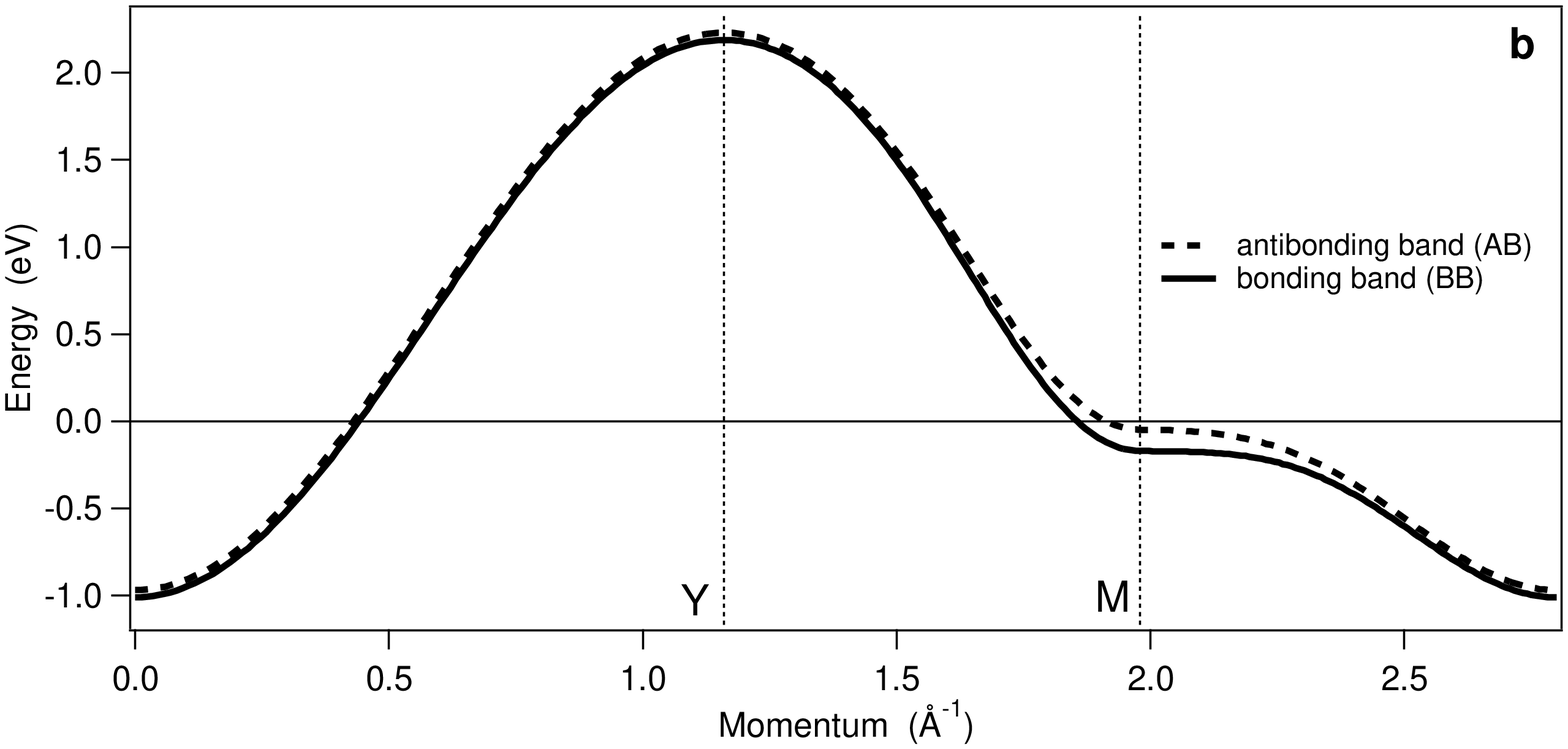}%
\includegraphics[width=2.972cm]{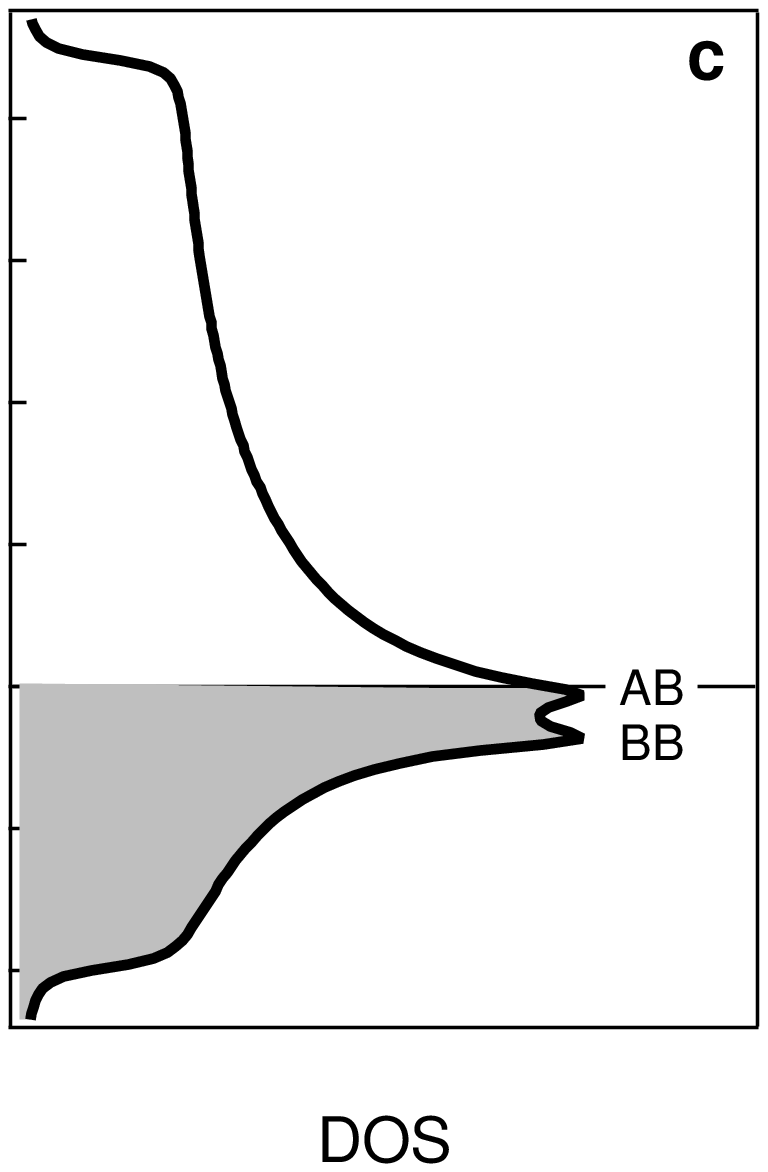}%
\caption{\label{Fig4} The bare band dispersion, as extracted from experiment \cite{20}: (a) Fermi surfaces, the dotted square denotes the boundary of he 1st BZ, the GYM triangle shows the path along which the dispersion is shown on panel (b); (c) corresponding density of states (DOS) with two vHs's close to $E_F$.}
\end{figure*}

\section{What is simple in HTSC?}

An idea of this section is to show that the cuprates, in terms of their electronic structure, appear to be not so unusual as it was believed before. This evolution from complexity to simplicity has been mainly a consequence of both the development of the ARPES technique (improving energy and momentum resolutions, statistics, increasing the excitation energy range) and accumulating experimental experience---increasing a signal-to-noise ratio on a large scale of our understanding. We believe that the important progress made in the last years consists in a conclusion that the cuprates in the doping range where superconductivity occurs are rather simple, in a sense that we can decompose the quasi­particle spectrum over the whole Brillouin zone into \textit{well predictable band structure} of the non-interacting electrons and the interaction effect in terms of \textit{quasiparticle self-energy}.

\textbf{HTSC band structure.} Earlier ARPES experiments had made a great deal of progress in our understanding of underlying electronic properties of HTSC cuprates but also, inevitably, were sometimes misleading giving rise to mistaken ideas. Such was the belief that the observed electronic structure of the cuprates cannot be appropriately described by the local-density approximation (LDA) band structure calculations \cite{24}. Among the features that could not be accounted for by the LDA were: extended flat bands \textquotedblleft indistinguishable" from $E_F$ \cite{22} that gave rise to ideas about \textquotedblleft pinning" of the van Hove singularity to the Fermi level \cite{25} and to a number of \textquotedblleft narrow band" descriptions \cite{26}; strong propensity for $(\pi, \pi)$ FS nesting \cite{22}; or experimental \textquotedblleft evidence" for the absence of the bilayer splitting (BS) \cite{27} that was taken as evidence for electronic confinement within the planes of CuO-bilayer due to strong correlations and de-confinement driven superconductivity \cite{28}, and had also many other consequences.

Now, it is well established that the hole-doped cuprates exhibit well defined \textquotedblleft large" FS \cite{23}, such as shown in Fig.~3. \textquotedblleft Large" means that the FS area (in holes) is equal to $(1 + x)/2$, where x is the hole density (or doping concentration). We have shown \cite{19} that such a relation holds for a wide doping range: $x = 0.1 - 0.2$. This means that the FS precisely satisfies the Luttinger theorem---the FS does not depend on interactions. This simple result has a far-reaching consequence allowing to conclude that \textit{superconductivity in cuprates emerges from a metallic phase} but not from the AF insulating phase, the evolution from which would require a \textquotedblleft small" $(\propto x)$ hole-like FS.\footnote{Here we should mention the idea of FS \textquotedblleft arcs" to which the large FS transforms in the pseudo-gap state and which evolves into four FS points with further underdoping \cite{29}. Our data on pseudo-gap function \cite{30} do not support this idea, although we admit that measuring the gap in photoemission is not straightforward \cite{20} and the problem of the pseudo-gap is still waiting for careful study.}

From high resolution FS maps measured for Bi-2212 in a wide doping range, taking into account the applicability of the Luttinger theorem, we have derived, within the tight-binding (TB) model, the bare band dispersion for this compound and found its good agreement with LDA prediction \cite{20}. Fig.~4 shows a result of such a procedure---the bare band dispersion for an overdoped Bi(Pb)-2212 ($x$ = 0.22). Recent observation of small bilayer splitting along the nodal direction of Bi-2212 can be considered as the most precise measure of the agreement between LDA and experiment \cite{31}: 0.012(1) \AA$^{-1}$ in momentum or 48(4) meV in energy (ARPES), 0.015 \AA$^{-1}$ or 50 meV (LDA).\footnote{We note that from TB fit to the FS shape one can extract a shape of the dispersion without the energy scale (reduced TB parameters) \cite{20}. In order to determine the scale, one needs to analyze self-consistently the renormalization effects \cite{16}.} 

The ability to resolve the BS for the cuprates with two adjacent CuO layers per unit cell is a distinguishing feature of new century ARPES \cite{8}. Now it has been observed not only for overdoped samples \cite{8} but also for optimally- and underdoped ones \cite{19, 31, 32}.\footnote{Despite this, there are still ARPES results discussed in favor of the absence of the BS for optimally- and underdoped Bi-2212 \cite{6}.} The existence of the BS not only refutes the existence of the mentioned electronic confinement \cite{28} but also questions some interpretations of a \textquotedblleft complex" lineshape in terms of many-body effects in a number of photoemission spectra. Of particular interest has been the famous peak-dip-hump (PDH) structure seen in $(\pi,0)$ region \cite{33} (see Fig.~5b). This line shape was widely believed to be the result of a single spectral function (see \cite{34} and references therein), the details of which are expected to reveal, at a fundamental level, the identity of the interactions involved in the generation and perpetuation of the superconducting state in these systems \cite{35, 36}. Using a wide range of excitation energies we have shown \cite{34} that the PDH structure mainly has a structural reason---is caused by the BS. In overdoped case the structural PDH is entirely dominating, while for the underdoped compounds, the interaction really produces a PDH-like structure on the $(\pi,0)$ spectrum (intrinsic PDH) for only one antibonding band \cite{12}. The magnitude of such a \textquotedblleft dip" is much lower than the structural one and highly depends on doping: vanishes to the overdoped side \cite{37}. Thus, the interactions really effect the electronic density of states in the whole antinodal region \cite{12, 37, 39} in the way proposed earlier \cite{35, 36} but their influence is much less than it was believed before and, more important, is highly doping dependent.

Another interesting \textquotedblleft structure-related" issue is the position of the van Hove singularity (vHs), which is crucial for quantitative approbation of any proposed HTSC mechanisms, especially so-called van Hove scenarios \cite{40}. Despite its importance, the evolution of the vHs with doping in Bi-2212 was not addressed in previous experiments.\footnote{The observation of the \textquotedblleft extended" saddle point at 19 meV had been reported for Y-124 \cite{40} but this value should be considered with caution because breaking of the chain layer during cleavage should substantially change the doping level of the topmost CuO layer in respect to the bulk. In case of Bi-2212 this point was also complicated by a \textquotedblleft complex" PDH lineshape below Tc and very wide spectrum for higher temperatures, so only position of the $(\pi, 0)$ EDC peak ($\approx$ 50 meV \cite{27}), not of the bare band, could be determined.} Now we can firmly say that for the doping range from $x$ = 0.11 to 0.22 \cite{19} the position of the bonding bare band in the saddle point, $\varepsilon_B$(M), changes from $\approx$ 260 meV to 190 meV, and for the antibonding band, $\varepsilon_A$(M) -- from 90 meV to 30 meV. This shift is consistent with the rigid band model \cite{20} with $d\varepsilon/dx \approx$ 0.5 eV, and one can speculate that the onset of superconductivity at $x$ = 0.27 is structure related---happens when the antibonding vHs crosses the Fermi level. This intriguing hypothesis, however, seems not to be valid for one-layer compounds.\footnote{A hole- to electron-like FS topological transition has been reported for Bi-2201 \cite{41} and LSCO \cite{42}, although for Bi-2201 this result is not confirmed \cite{43}.}

To summarize, the resolved electronic band structure of the cuprates keeps no mystery in the doping range where superconductivity occurs.\footnote{Even the \textquotedblleft shadow band", which one can see on FS maps of BSCO, and which was suggested to result from AFM correlations \cite{18}, has been shown to have structural origin \cite{44}.} In the normal state, even in the pseudo-gap region \cite{45}, it is in good agreement with LDA band structure calculations. In the superconducting state a $d$-wave gap opens \cite{8}, influencing the dispersion in BCS manner \cite{45, 46}. The electronic interactions appear in renormalization effects: shifting and smearing out the distribution of the electronic states in the momentum-energy space.

\textbf{Role of interaction.} The interactions in cuprates are anisotropic. In the end, this appears as a d-wave superconducting gap \cite{8}, but a strong anisotropy is also present in the normal state appearing as a pseudo-gap \cite{1} and anisotropic scattering rate \cite{47}.\footnote{The existence of the bilayer splitting naturally questions the conclusions of Ref.~\onlinecite{47}, although it seems that, similarly to the PDH story, some less pronounced anisotropy really exists.} In order to illustrate the role of interactions in HTSC, we consider the nodal direction, along which the both superconducting gap and pseudo-gap vanish. 

\begin{figure}[t]
\includegraphics[width=8cm]{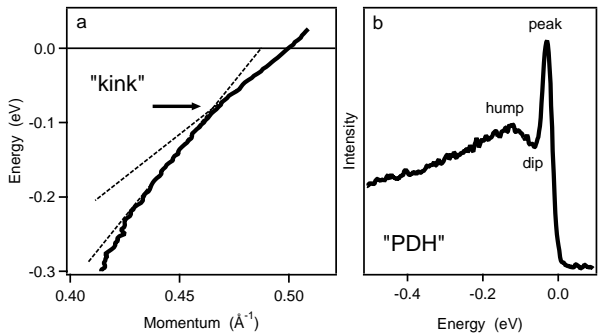}%
\caption{\label{Fig5} The best known representatives for the nodal and anti-nodal regions: (a) a \textquotedblleft 70 meV kink" on the renormalized dispersion \cite{16} and (b) a \textquotedblleft peak-dip-hump" (PDH) lineshape of a $(\pi,0)$ EDC \cite{12}.}
\end{figure}

\begin{figure*}[t]
\includegraphics[width=17.4cm]{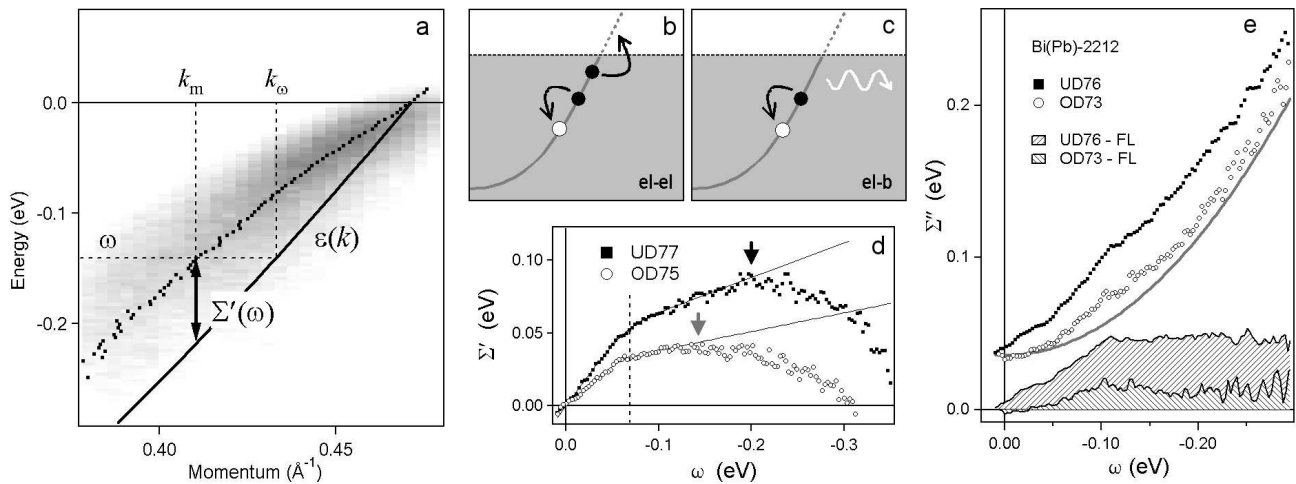}%
\caption{\label{Fig6} Nodal spectra analysis \cite{16}. (a) Bare band dispersion (solid line) and renormalized dispersion (points) on top of the spectral weight of interacting electrons (in gray scale). Though intended to be general, this sketch represents the nodal direction of an underdoped Bi-2212. The illustrations for the Auger-like scattering (b) and scattering by bosons (c). The real (d) and imaginary (e) parts of the self-energy for different doping levels. The solid line in (e) represents a Fermi liquid parabola---the Auger-like scattering.}
\end{figure*}

Fig.~6a represents essentials of the nodal spectra analysis. The real distribution of the quasiparticle spectral weight is shown as a blurred region in a grey scale. It differs from the bare dispersion of non-interacting electrons, $\varepsilon(k)$, in two ways: it is shifted and smeared out. These can be described by two quantities: mass renormalization and finite lifetime, which, in terms of Eq.~(1), are equal to the real and imaginary parts of the self-energy, respectively. In case of weak dependence on k, the real part is just the difference between the bare dispersion and renormalized dispersion (derived as a set of positions of MDC maxima), while the imaginary part $\Sigma"(\omega) = v_F W(\omega)$, where $W(\omega)$ is the half width at half maximum of this quasiparticle cloud (MDC HWHM) and $v_F$, in a simplest case of linear dispersion $\varepsilon(k) = v_F (k - k_F)$, is the bare Fermi velocity. Given that $\varepsilon(k)$ is \textit{a priori} unknown the self-energy parts cannot be extracted from the photoemission spectra unless an additional correlation between these quantities takes place. If the quasiparticle description is applicable to the cuprates, the $\Sigma(\omega)$, being derived from the causality-consistent Green's function, should be an analytical function, the real and imaginary parts of which are related by the Kramers-Kronig (KK) transformation \cite{48}. This transformation is, however, non-local, meaning that both $\Sigma(\omega)$ and $\varepsilon(k)$ cannot be disentangled from APRES spectra locally (in narrow energy range, in practice). Nevertheless, we have shown recently \cite{16} that, going deeper in energy, such a spectrum can be \textit{self-consistently} described by Eq.~(1) in a sense that the derived self-energy parts stay KK-related. This demonstrates the applicability of the quasiparticle approach to HTSC and opens a way to extract both the bare band structure and interaction functions from ARPES. Figs. 6d and 6e show some results of such a procedure: $\Sigma'(\omega)$ \cite{16} and $\Sigma''(\omega)$ \cite{49}. So, the main conclusion of this section is that the renormalization effect---the quasiparticle self-energy---is the only place where the mystery of HTSC is hidden. To resolve this mystery, one should just understand the nature of interactions which form the quasiparticles.

\section{Nature of interaction}

In order to understand the renormalization phenomenon in cuprates one should study its evolution on the phase diagram arena---as a function of doping and temperature. Here we focus on two key spots in the Brillouin zone: nodal and antinodal regions.

First we go over the nodal direction. Since 1999 \cite{17}, an important feature is seen by ARPES in the nodal spectra---a so-called \textquotedblleft kink" on the renormalized dispersion $\approx$ 70 meV below $E_F$ (see Fig.~5a and \cite{8, 9}). Since one does not expect such a feature on the bare dispersion here, neither of such an energy scale nor so sharp, it has been treated as a renormalization effect \cite{17}, and so it is. Therefore, a kink on the renormalized dispersion is a result of a kink on $\Sigma'(\omega)$ and, as a general consequence of causality, on $\Sigma''(\omega)$ \cite{49}. Historically, the kink has been associated first with a coupling to a bosonic mode \cite{50}, and the magnetic resonance observed by inelastic neutron scattering (INS) (see \cite{36} and references therein) has been suggested as a first candidate. Later \cite{51}, the same group has reported a ubiquity of the kink for a number of families of HTSC in a wide range of doping and temperature, and concluded in favour of phononic nature of the mode. It is the most important result of recent studies \cite{16, kinks, 49} that both kinks on the renormalization \cite{16, kinks} and scattering \cite{49} have appeared to be \textit{highly doping and temperature dependent}. The reason why in the previous studies such a strong dependence was not highlighted can be explained in part by unresolved bilayer splitting \cite{31} but mainly by influential contribution of strong electron-electron interaction (see Fig.~6d). 

In Ref.~\onlinecite{49}, we have managed to distinguish two scattering channels: the Auger-like decay and interaction with a mode. Panels (b) and (c) in Fig.~6 illustrate them, respectively. These two mechanisms describe the scattering of the photohole. In both the hole is filled by an overlying electron but in the former the energy excess is used to excite another electron above the Fermi-level (Auger-like process) while in the later it is taken off by a boson. For a simplest case of constant density of states (DOS), the first process gives a typical Fermi-liquid (FL) behavior---the quadratic dependence of the scattering rate on energy (or, for finite temperature, $\Sigma''(\omega) \propto \omega^2 + \pi^2 T^2)$. The bosonic scattering yields a convolution of the occupied DOS with the bosonic spectrum and, in case of a single bosonic mode, is a step-like function stepped at the mode energy. Fig.~6e shows the experimental scattering rate for an underdoped (UD) and overdoped (OD) Bi-2212 and its subdivision into two such channels. In reality, the Auger channel may deviate from a simple parabola due to peculiarities in the DOS, caused by van Hove singularities, gaps in the spectrum, etc., but these deviations have been found to be marginal \cite{49}. In contrast, the bosonic channel exhibits strong dependence on temperature and doping---it completely vanishes with overdoping and above $T$*. Therefore, we associate the bosonic spectrum in question with the spectrum of magnetic excitations (spin-fluctuations), which reveal similar dependencies on doping and temperature \cite{36}. In addition, very recently \cite{52, YBCO} we have found that the scattering between the bonding and antibonding bands exhibits an odd parity, similarly to the parity of spin-fluctuation spectrum.

Beyond the mentioned properties the magnetic spectrum exhibits a sophisticated structure over the momentum plane, peaked at $(\pi,\pi)$ (see \cite{36} and references therein). This turns us to the antinodal region, where in fact the renormalization effects have been found to be much stronger. Like the \textquotedblleft kink" for the nodal region, the heavily discussed PDH lineshape \cite{33} (see Fig.~5d and section III) is an undoubted logo of the antinodal area. It had been observed long before the kink but the story of its understanding is very similar. First it has been associated with the magnetic resonance but later, being observed in the whole doping range in Bi-2212 \cite{53}, it has been reinterpreted in terms of phonons \cite{54}. Later we have shown that, due to the bilayer splitting, one should distinguish a big \textquotedblleft structural" PDH \cite{34} and a much weaker \textquotedblleft intrinsic" PDH \cite{12}. The former is just a superposition of two bands, as it is seen in Fig.~2b for an OD sample. It makes the main contribution in PDH lineshape of the $(\pi,0)$ EDC over the whole doping range. The later is negligible for OD samples but increases with underdoping producing, in total, a clear formation of a \textquotedblleft gapped" region at a finite binding energy (manifestation of a sharp bosonic mode), as it is shown in Fig.~2d \cite{12}. Varying the excitation energy we have disentangled the effects of splitting and renormalization and evaluated the dependence of the coupling constant on doping. From this, one can conclude that (i) the sharp bosonic mode appears below $T_c$ \cite{12} and (ii) the scattering to it starts from about $x$ = 0.24 increasing with underdoping so that at $x$ = 0.12 it is about 4 times higher than the Auger-like electron-electron scattering (see Fig.~3 in \cite{37}).

To summarize this section, the interaction which dresses the quasiparticles in cuprates can be subdivided into two channels: a conventional electron-electron scattering due to Coulomb interaction through the Auger-like decay, and an electron-boson scattering. All of the properties of the later---doping, temperature and momentum dependence, as well as the parity---unambiguously point to its intimate relation with the spectrum of spin-fluctuations. Therefore, it is the interaction between electrons, via both Coulomb interaction and spin exchange, that makes an overwhelming contribution to quasiparticle dynamics in cuprates.  The coupling to phonons is much weaker and seems to be irrelevant for HTSC problem. 

\section{Conclusions and Outlooks}

In this short contribution we have reviewed the recent ARPES results which mark the way of evolution in our understanding of electronic structure of superconducting cuprates---a way from complexity to simplicity. We say \textquotedblleft simple" and \textquotedblleft complex" having in mind \textquotedblleft mainly understandable" and \textquotedblleft still mysterious".

\textit{So, what is simple in HTSC?} These are (i) the bare electronic structure and (ii) the way how the interactions appear. The \textit{band structure} seen by ARPES is very similar to one predicted by LDA calculations. Its evolution with doping well follows the rigid band approximation and suggests a structural mechanism of the onset of superconductivity at the overdoped side. The bilayer splitting for the compounds with two adjacent CuO layers per unit cell has essentially complicated situation earlier but now can be resolved and taken into account. Moreover, the splitting seems to be crucial for high $T_c$'s of these compounds---introducing a strong odd scattering channel. It is surprising that, after 20 years of extensive study, the \textit{dominant interactions} in cuprates have appeared to be describable by standard quasiparticle approach, in terms of the self-energy. Saying \textquotedblleft dominant", we have in mind the pseudo-gap formation which seems to fall out the quasiparticle approach, although, to make such a conclusion, careful study of the antinodal region is still needed. 

\textit{Now, what remains complex?} This is the mechanism of interaction. Although it is more or less clear that the spectrum of magnetic excitations, namely spin-fluctuations, forms the main interaction from which the pairing originates, the exact realization of such a mechanism still waits to be discovered. The progress is anticipated in two directions. First, it is ARPES measurements with an improved accuracy and their correlation with INS experiments. The fine structures of the pseudo-gap \cite{30} and superconducting gap \cite{55} should give the information about an exact momentum distribution of the scattering and coupling bosons, and one of the most anticipated result is to locate the real gap maximum---either it is in A-point or in \textquotedblleft hot-spots". Second is the location of the interaction in the real space, for which finding a correlation between ARPES and STS becomes of great importance \cite{56, 57, 58}. For instance, a hypothesis about phase separation into metallic and isolating regions should be checked and the role of magnetic excitations in the later should be clarified.

Finally, it seems that the way from complexity to simplicity is one which leads out of the HTSC labyrinth. Although the light at the end is already seen, the unification of different experimental techniques is needed to path it through.
  
We acknowledge the stimulating discussions with J. Fink, M. Knupfer, V. B. Zabolotnyy, M. S. Golden, A. N. Yaresko, S.-L. Drechsler, O. K. Andersen, I. Eremin, A. V. Chubukov, N. M. Plakida, Yu. N. Kucherenko, V. M. Loktev and many others. The project is part of the Forschergruppe FOR538.


\begin{thebibliography}{}

\bibitem{1} T. Timusk and B. Statt, \textit{Rep. Prog. Phys.} \textbf{62}, 61 (1999).

\bibitem{2} Y. Ando, S. Komiya, K. Segawa, S. Ono, and Y. Kurita, \textit{Phys. Rev. Lett.} \textbf{93}, 267001 (2004); cond-mat/0403032.

\bibitem{3} E. Dagotto, \textit{Rev. Mod. Phys.} \textbf{66}, 763 (1994).

\bibitem{4} V. J. Emery and S. A. Kivelson, \textit{Phys. Rev. Lett.} \textbf{74}, 3253 (1995).

\bibitem{5} A. Kampf and J. R. Schrieffer, \textit{Phys. Rev. B} \textbf{41}, 6399 (1990).

\bibitem{6} A. Kaminski, S. Rosenkranz, H. M. Fretwell, Z. Z. Li, H. Raffy, M. Randeria, M. R. Norman, and J. C. Campuzano, \textit{Phys. Rev. Lett.} \textbf{90}, 207003 (2003).

\bibitem{7} S. A. Kivelson, I. P. Bindloss, E. Fradkin, V. Oganesyan, J. M. Tranquada, A. Kapitulnik, and C. Howald, \textit{Rev. Mod. Phys.} \textbf{75}, 1201 (2003).

\bibitem{8} A. Damascelli, Z. Hussain, and Z.-X. Shen, \textit{Rev. Mod. Phys.} \textbf{75}, 473 (2003).

\bibitem{9} J. C. Campuzano, M. R. Norman, and M. Randeria, \textit{in Physics of Conventional and Unconventional Superconductors}, Vol. 2, edited by K. H. Bennemann and J. B. Ketterson, Springer-Verlag, Berlin, (2004).

\bibitem{10}    A. A. Abrikosov, L. P. Gor'kov, and I. E. Dzyaloshinskii, \textit{Quantum Field Theoretical Methods in Statistical Physics}, Pergamon, Oxford (1965).

\bibitem{11}    S. Hufner, Photoelectron Spectroscopy, \textit{ser. Solid State Sci.}, Vol. 82, Springer-Verlag, Berlin (1995).

\bibitem{12}    S. V. Borisenko, A. A. Kordyuk, T. K. Kim, A. Koitzsch, M. Knupfer, M. S. Golden, J. Fink, M. Eschrig, H. Berger, and R. Follath, \textit{Phys. Rev. Lett.} \textbf{90}, 207001 (2003); cond-mat/0209435.

\bibitem{13}    S. V. Borisenko, A. A. Kordyuk, S. Legner, C. Durr, M. Knupfer, M. S. Golden, J. Fink, K. A. Nenkov, D. Eckert, G. Yang, S. Abell, H. Berger, L. Forro, B. Liang, A. Maljuk, C. T. Lin, B. Keimer, \textit{Phys. Rev. B} \textbf{64}, 094513 (2001); cond-mat/0102323.

\bibitem{14}    A. Kaminski, S. Rosenkranz, H. M. Fretwell, J. Mesot, M. Randeria, J. C. Campuzano, M. R. Norman, Z. Z. Li, H. Raffy, T. Sato, T. Takahashi, and K. Kadowaki, \textit{Phys. Rev. B} \textbf{69}, 212509 (2004).

\bibitem{15}    S. V. Borisenko, T. Kim, A. A. Kordyuk, M. Knupfer, J. Fink, J. E. Gayone, Ph. Hofmann, H. Berger, B. Liang, A. Maljuk, and C. T. Lin, \textit{Physica C} \textbf{417}, 1 (2004). 

\bibitem{16}    A. A. Kordyuk, S. V. Borisenko, A. Koitzsch, J. Fink, M. Knupfer, and H. Berger, \textit{Phys. Rev. B} \textbf{71}, 214513 (2005); cond-mat/0405696; cond-mat/0510421.

\bibitem{17}    T. Valla, A. V. Fedorov, P. D. Johnson, B. O. Wells, S. L. Hulbert, Q. Li, G. D. Gu, and N. Koshizuka, \textit{Science} \textbf{285}, 2110 (1999).

\bibitem{18}    P. Aebi, J. Osterwalder, P. Schwaller, L. Schlapbach, M. Shimoda, T. Mochiku, and K. Kadowaki, \textit{Phys. Rev. Lett.} \textbf{72}, 2757 (1994). 

\bibitem{19}    A. A. Kordyuk, S. V. Borisenko, M. S. Golden, S. Legner, K. A. Nenkov, M. Knupfer, J. Fink, H. Berger, L. Forro, and R. Follath, \textit{Phys. Rev. B} \textbf{66}, 014502 (2002); cond-mat/0201485. 

\bibitem{20}    A. A. Kordyuk, S. V. Borisenko, M. Knupfer, and J. Fink, \textit{Phys. Rev. B} \textbf{67}, 064504 (2003); cond-mat/0208418. 

\bibitem{21}    A. A. Kordyuk, S. V. Borisenko, A. Koitzsch, J. Fink, M. Knupfer, B. Buechner, and H. Berger, \textit{J. Phys. Chem. Solids} (2005) to be published; cond-mat/0508574.

\bibitem{22}    D. S. Dessau, Z.-X. Shen, D. M. King, D. S. Marshall, L. W. Lombardo, P. H. Dickinson, A. G. Loeser, J. DiCarlo, C.-H. Park, A. Kapitulnik, and W. E. Spicer, \textit{Phys. Rev. Lett.} \textbf{71}, 2781 (1993).

\bibitem{23}    S. V. Borisenko, M. S. Golden, S. Legner, T. Pichler, C. Duerr, M. Knupfer, J. Fink, G. Yang, S. Abell, and H. Berger, \textit{Phys. Rev. Lett.} \textbf{84}, 4453 (2000); cond-mat/9912289.

\bibitem{24}    O. K. Andersen, A. I. Liechtenstein, O. Jepsen, and F. Paulsen, \textit{J. Phys. Chem. Solids} \textbf{56}, 1573 (1995).

\bibitem{kinks} A. A. Kordyuk, S. V. Borisenko, V. B. Zabolotnyy, J. Geck, M. Knupfer, J. Fink, B. Buechner, C. T. Lin, B. Keimer, H. Berger, Seiki Komiya, and Yoichi Ando, cond-mat/0510760.

\bibitem{YBCO} S. V. Borisenko, A. A. Kordyuk, V. Zabolotnyy, J. Geck, D. Inosov, A. Koitzsch, J. Fink, M. Knupfer, B. Buechner, V. Hinkov, C. T. Lin, B. Keimer, T. Wolf, S. G. Chiuzbaian, L. Patthey, and R. Follath, cond-mat/0511596.

\bibitem{25}    A. I. Liechtenstein, O. Gunnarsson, O. K. Andersen, and R. M. Martin, \textit{Phys. Rev. B} \textbf{54}, 12505 (1996).

\bibitem{26}    E. Dagotto, A. Nazarenko, and M. Boninsegni, \textit{Phys. Rev. Lett.} \textbf{73}, 728 (1994).

\bibitem{27}    H. Ding, A. F. Bellman, J. C. Campuzano, M. Randeria, M. R. Norman, T. Yokoya, T. Takahashi, H. Katayama-Yoshida, T. Mochiku, K. Kadowaki, G. Jennings, and G. P. Brivio, \textit{Phys. Rev. Lett.} \textbf{76}, 1533 (1996).

\bibitem{28}    P. W. Anderson, \textit{The Theory of Superconductivity in the High-Tc Cuprates}, Princeton University Press, Princeton (1997).

\bibitem{29}    M. R. Norman, H. Ding, M. Randeria, J. C. Campuzano, T. Yokoya, T. Takeuchik, T. Takahashi, T. Mochiku, K. Kadowaki, P. Guptasarma, and D. G. Hinks, \textit{Nature} \textbf{392}, 157 (1998).

\bibitem{30}    S. V. Borisenko, A. A. Kordyuk, A. Koitzsch, M. Knupfer, J. Fink, H. Berger, and C. T. Lin, \textit{Nature} \textbf{431} (02 September 2004); doi:10.1038/nature02931; cond-mat/0402251. 

\bibitem{31}    A. A. Kordyuk, S. V. Borisenko, A. N. Yaresko, S.-L. Drechsler, H. Rosner, T. K. Kim, A. Koitzsch, K. A. Nenkov, M. Knupfer, J. Fink, R. Follath, H. Berger, B. Keimer, S. Ono, and Yoichi Ando, \textit{Phys. Rev. B} \textbf{70}, 214525 (2004); cond-mat/0311137.

\bibitem{32}    Y.-D. Chuang, A. D. Gromko, A. V. Fedorov, Y. Aiura, K. Oka, Yoichi Ando, M. Lindroos, R. S. Markiewicz, A. Bansil, and D. S. Dessau, \textit{Phys. Rev. B} \textbf{69}, 094515 (2004).

\bibitem{33}    D. S. Dessau, B. O. Wells, Z.-X. Shen, W. E. Spicer, A. J. Arko, R. S. List, D. B. Mitzi, and A. Kapitulnik, \textit{Phys. Rev. Lett.} \textbf{66}, 2160 (1991).

\bibitem{34}    A. A. Kordyuk, S. V. Borisenko, T. K. Kim, K. A. Nenkov, M. Knupfer, J. Fink, M. S. Golden, H. Berger, and R. Follath, \textit{Phys. Rev. Lett.} \textbf{89}, 077003 (2002); cond-mat/0110379.

\bibitem{35}    M. R. Norman, H. Ding, J. C. Campuzano, T. Takeuchi, M. Randeria, T. Yokoya, T. Takahashi, T. Mochiku, and K. Kadowaki, \textit{Phys. Rev. Lett.} \textbf{79}, 3506 (1997).

\bibitem{36}    M. Eschrig and M. R. Norman, \textit{Phys. Rev. Lett.} \textbf{85}, 3261 (2000); \textit{Phys. Rev. B} \textbf{67}, 144503 (2003).

\bibitem{37}    T. K. Kim, A. A. Kordyuk, S. V. Borisenko, A. Koitzsch, M. Knupfer, H. Berger, and J. Fink, \textit{Phys. Rev. Lett.} \textbf{91}, 167002 (2003); cond-mat/0303422.

\bibitem{39}    A. D. Gromko, A. V. Fedorov, Y.-D. Chuang, J. D. Koralek, Y. Aiura, Y. Yamaguchi, K. Oka, Yoichi Ando, and D. S. Dessau, \textit{Phys. Rev. B} \textbf{68}, 174520 (2003); cond-mat/0205385.

\bibitem{40}    K. Gofron, J. C. Campuzano, A. A. Abrikosov, M. Lindroos, A. Bansil, H. Ding, D. Koelling, and B. Dabrowski, \textit{Phys. Rev. Lett.} \textbf{73}, 3302 (1994).

\bibitem{41}    T. Takeuchi, T. Yokoya, S. Shin, K. Jinno, M. Matsuura, T. Kondo, H. Ikuta and U. Mizutani, \textit{J. Electron Spectrosc. Relat. Phenom.} \textbf{114-116}, 629 (2001).

\bibitem{42}    A. Ino, C. Kim, M. Nakamura, T. Yoshida, T. Mizokawa, A. Fujimori, Z.-X. Shen, T. Kakeshita, H. Eisaki, and S. Uchida, \textit{Phys. Rev. B} \textbf{65}, 094504 (2002).

\bibitem{43}    T. Sato, T. Kamiyama, T. Takahashi, J. Mesot, A. Kaminski, J. C. Campuzano, H. M. Fretwell, T. Takeuchi, H. Ding, I. Chong, T. Terashima, and M. Takano, \textit{Phys. Rev. B} \textbf{64}, 054502 (2001).

\bibitem{44}    A. Koitzsch, S. V. Borisenko, A. A. Kordyuk, T. K. Kim, M. Knupfer, J. Fink, M. S. Golden, W. Koops, H. Berger, B. Keimer, C. T. Lin, S. Ono, Y. Ando, and R. Follath, \textit{Phys. Rev. B} \textbf{69}, 220505 (2004); cond-mat/0401114. 

\bibitem{45}    T. Eckl, W. Hanke, S. V. Borisenko, A. A. Kordyuk, T. Kim, A. Koitzsch, M. Knupfer, and J. Fink, \textit{Phys. Rev. B} \textbf{70}, 094522 (2004); cond-mat/0402340.

\bibitem{46}    J.C. Campuzano, H. Ding, M.R. Norman, M. Randeria, A.F. Bellman, T. Yokoya, T. Takahashi, H. Katayama-Yoshida, T. Mochiku, and K. Kadowaki, \textit{Phys. Rev. B} \textbf{53}, 14737 (1996).

\bibitem{47}    Z.-X. Shen and J. R. Schrieffer, \textit{Phys. Rev. Lett.} \textbf{78}, 1771 (1997).

\bibitem{48}    L. D. Landau and E. M. Lifschitz, \textit{Statistical Physics}, part 1, Pergamon, Oxford, (1980).

\bibitem{49}    A. A. Kordyuk, S. V. Borisenko, A. Koitzsch, J. Fink, M. Knupfer, B. Buechner, H. Berger, G. Margaritondo, C. T. Lin, B. Keimer, S. Ono, and Yoichi Ando, \textit{Phys. Rev. Lett.} \textbf{92}, 257006 (2004); cond-mat/0402643.

\bibitem{50}    P. V. Bogdanov, A. Lanzara, S. A. Kellar, X. J. Zhou, E. D. Lu, W. J. Zheng, G. Gu, J.-I. Shimoyama, K. Kishio, H. Ikeda, R. Yoshizaki, Z. Hussain, and Z. X. Shen, \textit{Phys. Rev. Lett.} \textbf{85}, 2581 (2000).

\bibitem{51}    A. Lanzara, P. V. Bogdanov, X. J. Zhou, S. A. Kellar, D. L. Feng, E. D. Lu, T. Yoshida, H. Eisaki, A. Fujimori, K. Kishio, J.-I. Shimoyama, T. Noda, S. Uchida, Z. Hussain, and Z.-X. Shen, \textit{Nature} \textbf{412}, 510 (2001).

\bibitem{52}     S. V. Borisenko, A. A. Kordyuk, A. Koitzsch, J. Fink, J. Geck, V. Zabolotnyy, M. Knupfer, B. Buechner, H. Berger, M. Falub, M. Shi, J. Krempasky, L. Patthey, cond-mat/0505192.

\bibitem{53}    D. L. Feng, D. H. Lu, K. M. Shen, C. Kim, H. Eisaki, A. Damascelli, R. Yoshizaki, J.-I. Shimoyama, K. Kishio, G. D. Gu, S. Oh, A. Andrus, J. O'Donnell, J. N. Eckstein, and Z.-X. Shen, \textit{Science} \textbf{289}, 277 (2000).

\bibitem{54}    T. Cuk, F. Baumberger, D. H. Lu, N. Ingle, X. J. Zhou, H. Eisaki, N. Kaneko, Z. Hussain, T. P. Devereaux, N. Nagaosa, and Z.-X. Shen, \textit{Phys. Rev. Lett.} \textbf{93}, 117003 (2004).

\bibitem{55}    S. V. Borisenko, A. A. Kordyuk, T. K. Kim, S. Legner, K. A. Nenkov, M. Knupfer, M. S. Golden, J. Fink, H. Berger, and R. Follath, \textit{Phys. Rev. B} \textbf{66}, 140509(R) (2002); cond-mat/0204557.

\bibitem{56}    B. W. Hoogenboom, C. Berthod, M. Peter, O. Fischer, and A. A. Kordyuk, \textit{Phys. Rev. B} \textbf{67}, 224502 (2003); cond-mat/0212329. 

\bibitem{57}    K. McElroy, R. W. Simmonds, J. E. Hoffmann, D.-H. Lee, J. Orenstein, H. Eisaki, S. Uchida, and J. C. Davis, \textit{Nature} \textbf{422}, 592 (2003). 

\bibitem{58}    A. A. Kordyuk, V. B. Zabolotnyy, D. S. Inosov, and S. V. Borisenko, cond-mat/0511638. 

\end{thebibliography}
\end{document}